\documentclass[10pt,letterpaper]{article}
\usepackage{spconf,amsmath,amssymb,graphicx,booktabs,array,etoolbox,capt-of,balance}
\usepackage[hidelinks]{hyperref}
\usepackage{url}
\usepackage{eso-pic}
\AtBeginDocument{%
  \AddToShipoutPictureFG*{%
    \AtPageLowerLeft{%
      \put(54,34){%
        \parbox[b]{504pt}{%
          \normalfont\fontsize{9}{11}\selectfont\raggedright
          This work has been submitted to the IEEE for possible publication.
          Copyright may be transferred without notice, after which this version
          may no longer be accessible.\par
        }%
      }%
    }%
  }%
}

\newcommand{\R}{\mathbb R}
\makeatletter
\patchcmd{\thebibliography}{\section{References}}{\section*{\centering\normalsize REFERENCES}}{}{\PackageError{ParaGeo}{Bibliography heading patch failed}{}}
\patchcmd{\thebibliography}{\usecounter{enumi}}{\usecounter{enumi}\setlength{\itemsep}{0.5pt}\setlength{\parsep}{0pt}}{}{\PackageError{ParaGeo}{Bibliography spacing patch failed}{}}
\patchcmd{\thebibliography}{\sloppy\clubpenalty4000\widowpenalty4000}{\sloppy\interlinepenalty10000\clubpenalty10000\widowpenalty10000}{}{\PackageError{ParaGeo}{Bibliography line-break patch failed}{}}
\makeatother
\title{PARAGEO: DECOMPOSING PARALINGUISTIC VARIATION INTO A SHARED LATENT GEOMETRY}
\name{Yuhan Liu$^{1}$, Yuxuan Ou$^{2}$, Ruoxi Su$^{3}$, Mohamed Ahmed Zaki$^{1}$, Yunbo Long$^{1}$ \sthanks{Corresponding author}}

\address{%
\small $^{1}$Department of Engineering, University of Cambridge\\
\small $^{2}$Department of Engineering Science, University of Oxford\\
\small $^{3}$Department of Computing \& Software Systems, University of Washington Bothell}

\begin{document}
\raggedbottom

\maketitle
\begin{abstract}

Speech delivery varies with both the requested paralinguistic attribute and the linguistic content. We introduce ParaGeo, a matched-content decomposition of paralinguistic variation in a frozen speech language model. Synthesized audio tokens are replayed with a fixed listening prompt; pooled key/value (K/V) representations are centered and projected into a shared low-dimensional space. Our GLM-4-Voice probe spans 80 requested controls from 12 benchmark families across eight sentences. With a globally fitted calibration basis, content-held-out centroid accuracy is 9.49\% versus a 1.25\% permutation baseline; same-label cross-content cosine similarity is 0.285 versus 0.017, and both conditional permutation tests yield $p = 0.001$. A separate ten-scenario, six-style probe reveals reproducible contrast directions across scenarios. Static, additive, and temporal interventions produce attribute-, layer-, and schedule-dependent response profiles. These results provide a shared coordinate representation for measuring paralinguistic structure and an empirical starting point for latent speech control.  Code is available at
\url{https://github.com/yuhanlydia/ParaGeo}.
\end{abstract}

\begin{keywords}
 paralinguistic representation, latent geometry, speech language models, speech control
\end{keywords}

\section{Introduction}
A sentence can be spoken with high pitch, slow pace, warm attitude, or hesitant delivery. Expressive speech therefore requires controlling delivery while preserving the lexical message. Learned style tokens, reference-conditioned prosody transfer, and information bottlenecks provide reusable speech factors~\cite{gst,prosodytransfer,speechsplit}. Semantic--acoustic token hierarchies~\cite{speechtokenizer} and speech-native dialogue models~\cite{glmvoice,moshi} bring this question to speech language models. We ask: \emph{which changes in vocal delivery recur across contexts, and how can heterogeneous attributes be described in a shared coordinate system?}

\begin{figure}[t]
\centering
\includegraphics[width=\columnwidth]{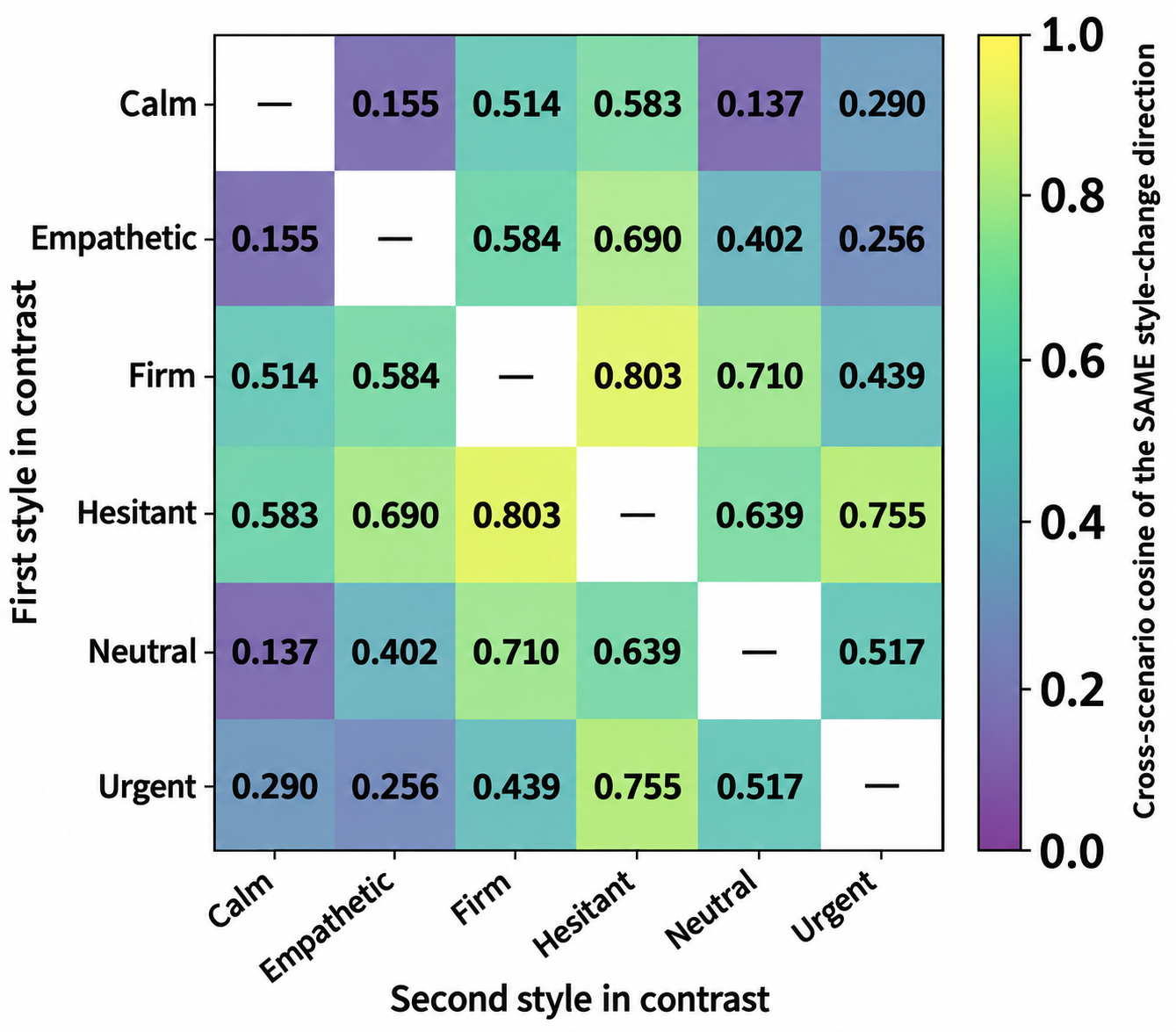}
\caption{\textbf{Vocal-style changes recur across contexts.}
Each cell measures the cross-scenario consistency of the same
style-difference direction, rather than similarity between styles.
Results use ten CRAD scenarios and two seeds.
All 15 unique contrasts pass Benjamini--Hochberg correction
at a false discovery rate of 0.05.}
\label{fig:discovery}
\end{figure}

In a ten-scenario debt-negotiation probe, six requested styles yield reproducible key/value (K/V) differences (Fig.~\ref{fig:discovery}). We therefore construct a broader probe of 80 controls from 12 SpeechParaling-Bench families~\cite{speechparaling}, crossed with eight common sentences. The resulting 16-dimensional coordinates yield a label-decoding accuracy of 9.49\% versus a permutation baseline of 1.25\%.

\begin{figure*}[t]
\centering
\includegraphics[width=\textwidth]{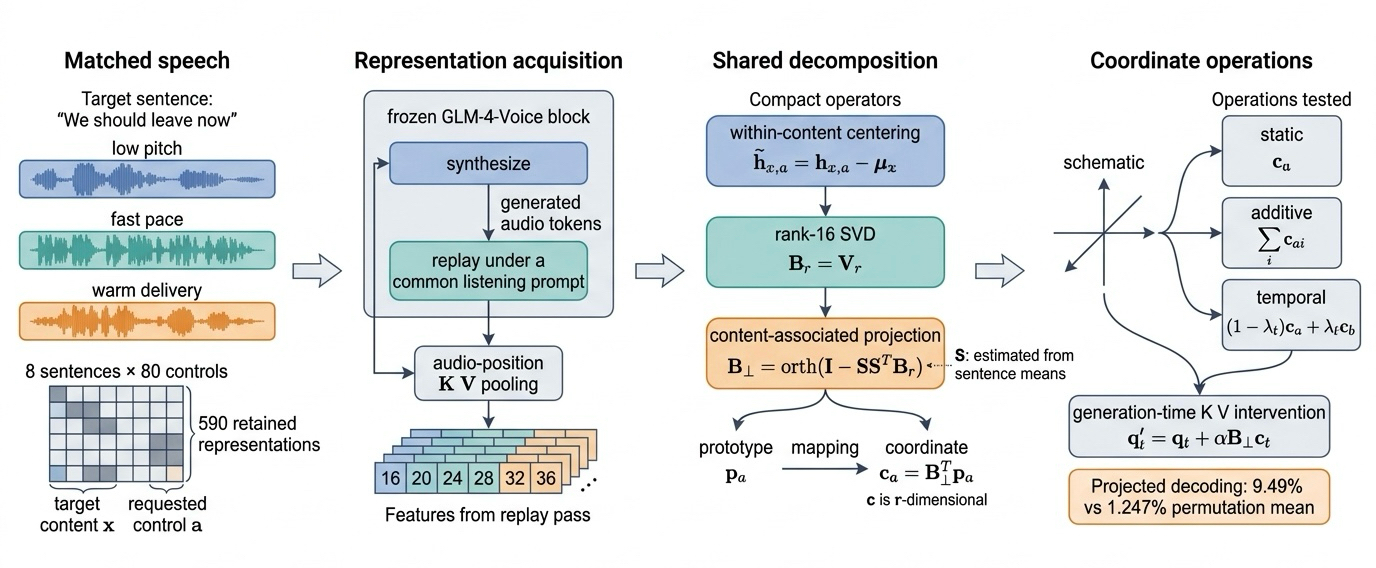}
\caption{\textbf{ParaGeo overview.} Matched-content renditions are synthesized and replayed through the frozen speech model. Audio-position K/V pooling, within-content centering, low-rank decomposition, and a content-associated projection yield shared coordinates. Static, additive, and temporal updates reuse the same basis. The diagram and waveforms are schematic.}
\label{fig:overview}
\end{figure*}
Activation Addition and Contrastive Activation Addition construct behavior directions~\cite{actadd,caa}; the linear representation hypothesis links probing and steering geometry~\cite{linearrep}. Expressive TTS methods either steer activations~\cite{emosteer,emoshift,cocoemo} or combine weight-space changes~\cite{emotionarithmetic,taskvector}. Emo-FiLM learns word-level modulation~\cite{emofilm}, while DisCo-Speech learns a content/prosody/timbre codec decomposition~\cite{discospeech}. CoCoEmo and subsequent geometric analyses examine emotion composition and steering sites~\cite{cocoemo,emotiongeometry}. We focus on a shared projection across heterogeneous request families in the replay-conditioned K/V representations of a frozen speech model.
\textbf{ParaGeo} contributes \textbf{(i)} cross-content evidence for requested paralinguistic structure, \textbf{(ii)} a frozen-model decomposition into shared low-dimensional coordinates, and \textbf{(iii)} measurements of alignment--discrimination and control--fidelity trade-offs. We compare static, additive, and temporal updates on the same fitted projection.

\section{ParaGeo}
\subsection{Matched-content acquisition and decomposition}
For a sentence $x$ and a requested attribute $a$, we synthesize one rendition, verify that its emitted text matches $x$, and replay its audio tokens through the same frozen model (Fig.~\ref{fig:overview}).
All replays use the same prompt: ``Listen to the speech and represent how it is spoken. Focus on paralinguistic delivery and ignore lexical content.''
We do not append the request label separately.
At the output of each selected fused QKV projection before attention, we average K and V separately over audio-token positions and concatenate the layerwise summaries into $h(x,a)\in\R^d$.

For the retained label set $\mathcal A_x$ of each sentence, define
\begin{equation}
\mu_x=\frac{1}{|\mathcal A_x|}\sum_{a\in\mathcal A_x}h(x,a),\qquad
\widetilde h(x,a)=h(x,a)-\mu_x.
\label{eq:center}
\end{equation}
Stacking the residuals yields $H\in\R^{N\times d}$.
A truncated SVD of $H$ gives $H\approx U_r\Sigma_rV_r^\top$, with $B_r=V_r\in\R^{d\times r}$.
Centering removes additive sentence offsets, and the SVD captures shared residual variation without preference labels or speech-model training.

\subsection{Shared projection and coordinate operations}
We separately obtain an orthonormal reference $S$ from an SVD of the centered sentence means.
We residualize $B_r$ against $S$ and average the residuals for each label:
\begin{equation}
B_\perp=\operatorname{orth}[(I-SS^\top)B_r],\quad
p_a=\frac{1}{|\mathcal X_a|}\sum_{x\in\mathcal X_a}\widetilde h(x,a),
\label{eq:basis}
\end{equation}
where $\mathcal X_a$ denotes the retained sentences for label $a$.
Then
\begin{equation}
c_a=B_\perp^\top p_a,\qquad d_a=B_\perp c_a=P_\perp p_a.
\label{eq:coordinates}
\end{equation}
Here $P_\perp=B_\perp B_\perp^\top$.
Unlike target-supervised removal in INLP or LEACE~\cite{inlp,leace}, $S$ is estimated from sentence means rather than labels.
The resulting projection provides a measurable content proxy; representation similarity and intervention effects are distinct quantities~\cite{amnesic,targeted_erasure}.

For concatenated generation-time K/V projections $q_t$, static steering adds $\alpha d_a$ to $q_t$.
For a combination $\mathcal C$, we use
\begin{equation}
q_t'=q_t+\alpha B_\perp\sum_{a\in\mathcal C}c_a
     =q_t+\alpha P_\perp\sum_{a\in\mathcal C}p_a.
\label{eq:composition}
\end{equation}
Thus, coordinate addition applies the common projector to the sum of independently calibrated prototypes.
For a transition $a\to b$, we use
\begin{equation}
c_t=(1-\lambda_t)c_a+\lambda_tc_b,\qquad
q_t'=q_t+\alpha B_\perp c_t.
\label{eq:dynamic}
\end{equation}
The gradual coefficient is $\lambda_t=\min(t/191,1)$; the sudden schedule switches at pass 96.
After 192 forward passes, the endpoint is held.
The index follows interleaved text/audio generation rather than waveform time.

\section{Experimental Setup}
\textbf{Model and calibration.} We use GLM-4-Voice-9B~\cite{glmvoice} with int4/NF4 weights, temperature 0.8, top-$p$ 0.8, and at most 768 generated tokens.
Our probe crosses 80 benchmark-derived requests with eight fixed sentences, yielding 640 requested pairs and 590 retained representations after text matching and audio-token availability checks.
This is a requested-label probe rather than an official recognition benchmark.
Features are extracted from layers $\{16,20,24,28,32,36\}$ with rank 16.
The content reference retains eight SVD columns, although eight centered means span at most seven nonzero-variance directions.

\textbf{Representation tests.} A cosine nearest-centroid classifier predicts one of 80 exact labels, using seven sentences to form centroids and the remaining sentence for testing.
The basis and unlabeled within-sentence centering use all calibration features, making this a fixed-basis, content-held-out centroid probe~\cite{probecontrols}.
We also measure same-label cross-content cosine similarity and raw-SVD explained variance.
For 1,000 conditional permutations, labels are independently shuffled within each sentence; $p=(1+b)/1001$, where $b$ is the number of null statistics at least as large as the observed value~\cite{permutation}.
The separate six-style discovery uses all 126 overlapping balanced 5/5 scenario splits, bidirectional evaluation, and Benjamini--Hochberg FDR across 15 style contrasts~\cite{bh}; these splits are not independent replications.

\textbf{Generation and controls.} Catalog-covered English SpeechParaling inputs use the waveform path and model-based pairwise evaluation protocol of SpeechParaling.
Development uses positions $0,5,10,\ldots$ in the eligible-item list; all other positions are held out.
We select $\alpha\in\{0.25,0.5,1,1.5\}$ by maximizing development preference subject to a prespecified two-percentage-point text-WER tolerance ($\Delta\mathrm{WER}\le0.02$), breaking ties toward smaller scales.
Static steering selects $\alpha=0.5$.
Composition uses $\alpha=0.25$ exploratorily because no scale satisfies the tolerance; dynamic steering uses $\alpha=0.25$ exploratorily because the selected scale 1.5 is incomplete.
Evaluation sets contain 168 static, 38 composition, and 60 dynamic items; ablations use the first 24 held-out items in benchmark-index order.

The reference is matched prompt-only generation.
Full-space composition sums the same prototypes and matches the global norm of the projected vector.
A Gaussian unit direction provides norm-matched controls; its orientation is fixed per dynamic item, while its norm follows the target schedule.
Judgments are scored as 1 for a win, 0.5 for a tie, and 0 for a loss; scores are first averaged within an item and then across items.
A preference score of 50 denotes parity.
Nominal 95\% intervals are computed using 10,000 item-level bootstrap resamples~\cite{significance}.
A balanced extension reuses 180 ParaGeo outputs (18 labels, ten inputs each) against five direction-control seeds; these are not 900 independent inputs.

\begin{figure*}[t]
\centering
\includegraphics[width=\textwidth]{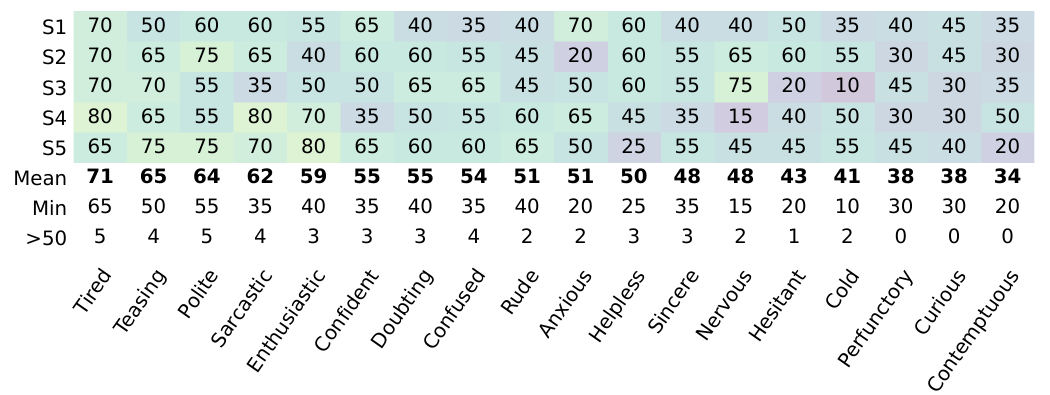}
\caption{\textbf{Attribute-specific responses across five direction controls.} Columns show all 18 attributes, ordered by mean; S1--S5 denote the five random-direction comparison seeds. Each column reuses ten inputs and the same ParaGeo outputs. Cells report direct preference; summaries are mean, minimum, and count above 50. The pooled preference is 51.50 [45.39,57.56].}
\label{fig:attribute_response}
\end{figure*}

\section{Results and Analysis}

\subsection{Cross-context structure and compact representation}
The discovery study (Fig.~\ref{fig:discovery}) yields a median six-style accuracy of 85.0\% with audio-token averaging, compared with 52.92\% for the last audio token and 16.7\% chance. All 15 style contrasts pass Benjamini--Hochberg FDR at 0.05. Mean cross-scenario consistency is 0.498; firm--hesitant reaches 0.803, compared with 0.137 for calm--neutral.
The repeatability of the same style change motivates both pooled acquisition and shared-coordinate analysis.

\begin{table}[t]\centering\small
\caption{\textbf{Representation measurements on 590 records.} Both rows are centered within sentence. Accuracy predicts the exact requested label; cosine compares the same label across sentences.}
\label{tab:geometry}
\vspace{3pt}
\setlength{\tabcolsep}{4pt}
\begin{tabular}{@{}lrr@{}}\toprule
Representation & 80-way accuracy & Cosine\\\midrule
Centered full K/V & 11.69\% & 0.152\\
Projected 16-D coordinates & 9.49\% & 0.285\\\bottomrule
\end{tabular}
\end{table}

Table~\ref{tab:geometry} separates discrimination from alignment. Projected decoding gives 56 correct labels, versus 69 in full space. Accuracy is 9.49\% against a permutation mean of 1.247\%; cosine is 0.285 against 0.017. Both tests yield $p=1/1001$. The raw rank-16 SVD captures 61.76\% of residual variance. Projection raises same-label cosine by a factor of 1.88 while reducing exact-label accuracy by 2.20 points: agreement within labels and separation between labels measure different properties of the shared geometry.

\subsection{Layer, rank, and schedule responses}
Table~\ref{tab:dynamic_tradeoff} compares all 11 dynamic configurations using the same control and fidelity metrics.
Middle-layer steering achieves the highest preference estimate, 68.75 [56.25,81.25], with $\Delta$WER $=-0.0060$.
Static middle-layer preference is 64.58 [50.00,79.17], compared with 60.42 for all layers.
Layer groups are defined as early $\{16,20\}$, middle $\{24,28\}$, and late $\{32,36\}$.
Because masking also changes perturbation energy, these results are exploratory operating configurations rather than an energy-matched localization test.

\begin{table}[t]\centering\small
\caption{\textbf{Dynamic control and text fidelity.} Fixed 24-item subset; preference against prompt-only (50 = parity), nominal 95\% bootstrap CIs, and absolute change in text-channel WER.}
\label{tab:dynamic_tradeoff}
\vspace{3pt}
\setlength{\tabcolsep}{3pt}\renewcommand{\arraystretch}{1.04}
\begin{tabular}{@{}lccc@{}}\toprule
Configuration & Pref.$\uparrow$ & 95\% CI & $\Delta$WER$\downarrow$\\\midrule
All layers ($r=16$)&54.17&[37.50,70.83]&$+0.0055$\\
w/o content proj.&43.75&[27.08,60.42]&$+0.0021$\\\midrule
Rank 4&58.33&[41.67,75.00]&$0.0000$\\
Rank 8&54.17&[37.50,70.83]&$+0.0108$\\
Rank 32&47.92&[31.25,64.58]&$+0.0021$\\\midrule
Early layers&60.42&[45.83,75.00]&$+0.0160$\\
Middle layers&\textbf{68.75}&[56.25,81.25]&$\mathbf{-0.0060}$\\
Late layers&60.42&[45.83,75.00]&$0.0000$\\\midrule
Start only&56.25&[39.58,70.83]&$+0.0010$\\
End only&47.92&[31.25,66.67]&$+0.0327$\\
Midpoint only&45.83&[29.17,62.50]&$0.0000$\\\bottomrule
\end{tabular}
\end{table}

Dynamic preference is nonmonotone in rank.
The all-layer schedule scores 54.17, compared with 56.25 for the static start coordinate.
These nominal comparisons separate the recovered representation from its injection operator.

\subsection{Attribute-specific response and projection trade-offs}
The complete balanced matrix (Fig.~\ref{fig:attribute_response}) exposes substantial heterogeneity. Tired and polite average 71 and 64 against the five controls, with minima of 65 and 55. Teasing averages 65, with four above-parity cells and one tie. All 18 attributes are shown: the pooled direct preference is 51.50 [45.39,57.56], and three of five seed-level estimates exceed 50. The highlighted labels are descriptive patterns, not independent confirmations of a general control gain. On 38 exploratory composition inputs, norm-matched low-rank updates yield 45.67\% text-channel WER, which is 4.61 points below full-space addition (50.28\%) but 6.88 points above prompt-only (38.79\%).
Direct preference for the projection over full-space addition is 42.98 [34.87,51.10]. Thus, the reduced text disturbance of the projection comes with a measurable control--fidelity trade-off. At the full-task level, static ParaGeo scores 55.36 [49.11,61.61] against prompt-only, while the direction control scores 55.95. Exploratory composition and dynamic preferences are 45.39 [36.18,54.82] and 51.67 [40.00,63.33], respectively.
Table~\ref{tab:dynamic_families} reports the six families in the 60-item dynamic set, which is distinct from the 18-label static extension.
Pace achieves a preference score of 75 over 12 inputs without family-specific tuning.

\noindent\begin{minipage}{\columnwidth}
\centering\small
\captionof{table}{\textbf{Dynamic family profiles.} All 60 inputs at exploratory $\alpha=0.25$; preference against prompt-only.}
\label{tab:dynamic_families}
\vspace{3pt}
\setlength{\tabcolsep}{9pt}
\begin{tabular}{@{}lrr@{}}\toprule
Family & Inputs & Preference\\\midrule
Emotion &12&54.17\\
Pace &12&75.00\\
Pitch &12&45.83\\
Rhythm &12&37.50\\
Timbre &4&62.50\\
Volume &8&37.50\\\bottomrule
\end{tabular}
\end{minipage}
\medskip

\section{Conclusion}

We identified recurring paralinguistic structure in GLM-4-Voice and introduced ParaGeo to represent it in shared coordinates. Cross-scenario contrasts, fixed-basis centroid probes, and conditional permutation tests support this structure. The projection increases same-label alignment while reducing fine-grained discrimination; generation tests reveal attribute-, layer-, and schedule-dependent responses. A shared basis makes vocal delivery an explicit object for comparison and editing, as attributes, combinations, and temporal paths reuse the same coordinates without updating the model. This provides an empirical starting point for latent speech control by separating the representation of delivery from its generation-time application. Future work will target training-only basis fitting, cross-model replication, and listener validation.

\clearpage
\balance
\begingroup
\small
\bibliographystyle{IEEEbib}
\bibliography{references}
\endgroup

\section{Funding Acknowledgments}
No funding was received for conducting this study. The authors have no relevant financial or nonfinancial interests to disclose.

\section{Compliance with Ethical Standards}
This is a study using public datasets for which no ethical approval was required.

\clearpage
\nobalance

\appendix
\setcounter{table}{0}
\setcounter{figure}{0}
\setcounter{equation}{0}
\renewcommand{\thetable}{A\arabic{table}}
\renewcommand{\thefigure}{A\arabic{figure}}
\renewcommand{\theequation}{A\arabic{equation}}
\renewcommand{\theHtable}{appendix.\arabic{table}}
\renewcommand{\theHfigure}{appendix.\arabic{figure}}
\renewcommand{\theHequation}{appendix.\arabic{equation}}
\pagestyle{plain}
\setlength{\parskip}{3pt plus 1pt}
\renewcommand{\arraystretch}{1.12}
\newenvironment{apptable}{\begin{table}[!t]\centering\small}{\end{table}}
\newenvironment{apptable*}{\begin{table*}[!t]\centering\small}{\end{table*}}
\newenvironment{appfigure}{\begin{figure}[!t]\centering}{\end{figure}}
\newenvironment{appfigure*}{\begin{figure*}[!t]\centering}{\end{figure*}}
\setcounter{topnumber}{3}
\setcounter{dbltopnumber}{2}

\section{Extended Results}
\label{app:evidence}

\subsection{A shared representation with measured intervention responses}
ParaGeo makes a heterogeneous set of vocal requests comparable within one frozen speech model. The central object is a common projection of replay-conditioned K/V representations. An attribute, a combination of attributes, and a temporal transition can all be expressed with that projection. The extended results answer two connected questions: whether requested delivery has recurring structure across lexical contexts, and how updates constructed from that structure affect generation. Keeping those questions explicit makes the favorable results interpretable without requiring every intervention to improve the aggregate benchmark score.

The strongest representation evidence is the conjunction of cross-content decoding, same-label alignment, and a conditional permutation control. In the 590-record calibration collection, projected exact-label accuracy is 9.49\%, compared with a permutation mean of 1.25\%; same-label cosine is 0.285, compared with 0.017. Both finite permutation estimates are $1/1001\approx0.001$. The raw rank-16 SVD captures 61.76\% of residual variance. These measurements support a compact, shared description of requested delivery under the acquisition protocol. The final projection also changes which distinctions are retained: same-label alignment increases while exact-label decoding decreases relative to centered full K/V.

The generation results locate useful responses within that representation. Middle-layer dynamic injection has a preference estimate of 68.75 on the fixed diagnostic subset, with a nominal interval of [56.25,81.25] and a text-WER change of $-0.0060$. In the balanced static study, tired and polite have means of 71 and 64 against five direction controls, and both remain above parity in every control column. Pace transitions score 75 on the 12 corresponding dynamic inputs. Each finding identifies an observed setting in which the coordinates produce a favorable response. Their full comparison sets, sampling units, and uncertainty appear below.

The aggregate results set the scope of those responses. The direct static comparison against five random-direction controls is 51.50 [45.39,57.56]. Composition reduces text disturbance relative to norm-matched full-space addition, but does not win the direct preference comparison. The complete dynamic evaluation uses an exploratory scale because the development-selected arm is incomplete. The resulting scientific account is a shared-coordinate representation with attribute- and operator-dependent response profiles; the evidence does not establish an overall controller advantage over the qualified direction controls.

\begin{apptable*}
\caption{\textbf{How to read the evidence.} Each row names a distinct measurement obligation. Preference uses a 0--100 scale with 50 denoting parity; differences in WER elsewhere are absolute fractions.}
\label{tab:evidence_map}
\setlength{\tabcolsep}{5pt}
\begin{tabular}{@{}p{.20\textwidth}p{.39\textwidth}p{.35\textwidth}@{}}\toprule
Question & Evidence in this record & Interpretation\\\midrule
Does a requested change recur across content? & Six-style cross-scenario contrasts; 80-way content-held-out centroids; conditional permutations & Request-associated structure in a specified replay representation.\\
What does the projection retain? & Raw-SVD variance; full versus projected decoding and cosine & Increased alignment with a loss of some fine-grained label distinctions.\\
Where do updates respond favorably? & Complete rank/layer/schedule ablations; 18-attribute control matrix; task-family profiles & Descriptive operating settings, with nominal subset intervals and complete unfavorable entries.\\
Is there an aggregate control gain? & Development selection; full held-out results; direct random and full-space comparisons & The current aggregate comparisons do not establish general superiority.\\\bottomrule
\end{tabular}
\end{apptable*}

\subsection{Discovery in spoken interaction}
The initial discovery collection contains ten debt-negotiation scenarios, two generation seeds, and six vocal styles: 120 unique representations. Across all 126 overlapping balanced 5/5 scenario partitions, evaluated in both directions, audio-token mean pooling gives a median six-way accuracy of 85.0\%, compared with 52.92\% for the last audio token and 16.7\% uniform chance. Every partition exceeds its reported shuffled 95th-percentile reference. All 15 oriented style contrasts pass the reported Benjamini--Hochberg procedure at FDR 0.05~\cite{bh}. Their mean cross-scenario direction consistency is 0.498; firm--hesitant reaches 0.803, whereas calm--neutral is 0.137.

These measurements motivate pooled acquisition: the same requested change can recur across scenarios even when lexical content changes. Contrast consistency is the cosine between the \emph{same oriented style difference} in different scenarios, rather than similarity between the two named styles. The repeated partitions reuse utterances and are not independent replications. Six-way scenario decoding and 80-way sentence decoding use different label spaces and preprocessing. Historical negotiation search did not establish a reliable terminal-utility gain; it supplies the discovery context, rather than a downstream success claim for ParaGeo. The complete contrast matrix, repeated-split confusion counts, and historical task summaries remain in the supplied data.

\section{Calibration and Representation Construction}
\label{app:acquisition}

\subsection{The matched-content collection}
The intended calibration design crosses eight fixed English sentences with 80 requested controls, giving 640 requested pairs. One rendition is generated per cell. A record is retained when the normalized emitted text matches the normalized target sentence and audio tokens are available for replay. The resulting matrix contains 590 rows. The aggregate record does not identify the individual reasons for all 50 omissions, so their causes are left unspecified. Table~\ref{tab:sentences} lists the complete lexical bank, and Table~\ref{tab:catalog} lists the actual control strings.

\begin{apptable}
\caption{\textbf{The eight fixed calibration sentences.} All controls reuse this lexical bank. The identifiers match the released calibration-sentence file.}
\label{tab:sentences}
\setlength{\tabcolsep}{3pt}
\begin{tabular}{@{}lp{.78\linewidth}@{}}\toprule
ID & Target text\\\midrule
\texttt{cal:00} & The morning train arrived exactly on time.\\
\texttt{cal:01} & Please place the blue notebook beside the lamp.\\
\texttt{cal:02} & We will review the final report after lunch.\\
\texttt{cal:03} & A small bird landed quietly on the garden fence.\\
\texttt{cal:04} & The museum opens at nine o'clock tomorrow.\\
\texttt{cal:05} & I left the keys on the wooden table near the door.\\
\texttt{cal:06} & The team finished the experiment before sunset.\\
\texttt{cal:07} & She carried a warm cup of tea into the study.\\\bottomrule
\end{tabular}
\end{apptable}

Synthesis uses a deterministic assignment indexed by content and lexicographically sorted control labels. Matching is checked in the model's emitted text channel. It is a lexical acquisition check, not independent waveform transcription or human confirmation of the requested style. In particular, the subsequent classification target is the exact requested label, such as \texttt{Pitch::high pitch}; it is neither a 12-family classification problem nor a human-annotated recognition benchmark.

The catalog includes utterance-level requests and word-local instructions. A request to stress ``must,'' for example, can be paired with a calibration sentence that does not contain that word. Such entries remain visible in the inventory because they were part of the actual design. They should be read as requested controls, rather than as 80 independently validated and mutually exclusive acoustic factors. Emotion, pitch, pace, and rhythm can covary. A common subspace does not assign an independent causal axis to each family.

\subsection{Synthesis followed by common-prompt replay}
Acquisition has two passes through frozen GLM-4-Voice~\cite{glmvoice}. The first synthesizes the target sentence under the control request and saves its audio-token IDs. The second replays those IDs with the same instruction for every control:
\begin{quote}
Listen to the speech and represent how it is spoken. Focus on paralinguistic delivery and ignore lexical content.
\end{quote}
The model's assistant streaming-transcription prefix is used in replay. No attribute name is separately appended to this prompt. Request information can therefore enter the measured representation through the synthesized tokens, including duration and other correlated properties of the rendition. This is a representation of listening to the output; it is not the original generation-state activation.

The recorder takes K and V at the output of the fused QKV projection, before attention. The sampled layers are $\{16,\allowbreak20,\allowbreak24,\allowbreak28,\allowbreak32,\allowbreak36\}$. For audio positions $\mathcal T_{\rm audio}$,
\begin{equation}
\begin{aligned}
\bar K^{(\ell)}&=\frac{1}{|\mathcal T_{\rm audio}|}\sum_{t\in\mathcal T_{\rm audio}}K_t^{(\ell)},\\
\bar V^{(\ell)}&=\frac{1}{|\mathcal T_{\rm audio}|}\sum_{t\in\mathcal T_{\rm audio}}V_t^{(\ell)}.
\end{aligned}
\end{equation}
The feature $h(x,a)$ concatenates these pooled chunks in the stored layer/K/V order. With $L$ layers and flattened K/V width $d_{kv}$, its dimension is $d=2Ld_{kv}$. The exact width follows the model configuration and artifact metadata. Branch IDs align features with calibration records; their association is not inferred from a common row order.

\begin{apptable*}
\caption{\textbf{Complete calibration catalog: 80 controls in 12 benchmark families.} Semicolons separate entries. The original benchmark wording, including ``loudly volume'' and distinct stress phrasings, is preserved.}
\label{tab:catalog}
\setlength{\tabcolsep}{5pt}
\begin{tabular}{@{}p{.17\textwidth}rp{.74\textwidth}@{}}\toprule
Family & Count & Requested controls\\\midrule
Age &4& child's voice; youthful voice; adult voice; elderly voice.\\
Pitch &5& very high pitch; high pitch; medium pitch; low pitch; very low pitch.\\
Timbre &6& bright timbre; hoarse timbre; smooth timbre; rich timbre; gentle timbre; sweet timbre.\\
Pace &5& very fast pace; fast pace; medium pace; slow pace; very slow pace.\\
Volume &5& shouting manner; loudly volume; normal volume; quietly volume; whisper.\\
Pause &3& clear pause after the word `listen'; clear pause after the word `but'; clear pause after the word `good'.\\
Rhythm &7& steady rhythm; lighthearted rhythm; soothing rhythm; rushed rhythm; emphatic rhythm; dragging rhythm; halting rhythm.\\
Stress &12& emphasis on `must'; emphasis on `is'; emphasis on `here'; stress on `must'; stress on `every day'; stress on `key'; heavy stress on `not allowed'; heavy stress on `very'; heavy stress on `must'; forceful tone on `now'; forceful tone on `careful'; forceful tone on `remember'.\\
Emotion &7& neutral emotion; happy emotion; sad emotion; angry emotion; surprised emotion; disgusted emotion; fearful emotion.\\
Cognitive State &9& confident tone; hesitant tone; confused tone; doubting tone; tired tone; curious tone; anxious tone; helpless tone; nervous tone.\\
Non-Linguistic Vocalizations &8& laughter; crying; sigh; coughing; scream; hiccups; yawn; smack of the lips.\\
Attitude &9& polite tone; sincere tone; enthusiastic tone; cold tone; sarcastic tone; contemptuous tone; rude tone; perfunctory tone; teasing tone.\\\midrule
Total &80& Each prototype is calibrated from single-control requests.\\\bottomrule
\end{tabular}
\end{apptable*}

\subsection{Centering, SVD, and the content proxy}
For sentence $x$, let $\mathcal A_x$ be its retained controls. Centering uses their mean $\mu_x$, as in Eq.~\ref{eq:center}. Writing $H_0$ for the uncentered feature matrix, the same operation is $H=CH_0$, with
\begin{equation}
\begin{aligned}
C&=\operatorname{blockdiag}(C_x),\\
C_x&=I_{n_x}-\frac{1}{n_x}\boldsymbol1\boldsymbol1^\top,\\
C_x^2&=C_x=C_x^\top.
\end{aligned}
\end{equation}
Consequently $C_x(H_x+\boldsymbol1u_x^\top)=C_xH_x$: an additive sentence offset is removed exactly. Under a balanced additive model $h(x,a)=u_x+v_a+\epsilon_{x,a}$, this leaves $v_a-\bar v+\epsilon_{x,a}-\bar\epsilon_x$. In the observed incomplete collection, the sentence-specific attribute mix and content--attribute interactions can also contribute to the residual. Centering provides a controlled comparison, rather than a guarantee of full linguistic invariance.

The raw right-singular vectors $B_r$ come from the full residual matrix. A separate SVD of centered sentence means supplies an orthonormal content proxy $S$. The implementation requests eight columns, although eight centered means have at most seven nonzero-variance directions and the helper does not truncate them by a singular-value tolerance. The eighth configured column is therefore not an established semantic factor. Residualizing $B_r$ gives $R=(I-SS^\top)B_r$; QR orthogonalization retains columns whose corresponding diagonal magnitude exceeds $10^{-9}$. The default stored basis has 16 columns.

The final coordinates are recalculated from the residual prototypes: $c_a=B_\perp^\top p_a$, $d_a=B_\perp c_a$. This sentence-mean construction differs from target-supervised removal in INLP and LEACE~\cite{inlp,leace}. It defines a reproducible content-associated projection that can be ablated; it does not certify removal of every lexical signal.

\section{Geometry Diagnostics and Their Interpretation}
\label{app:geometry}

\subsection{What the content hold-out tests}
In fold $x$, cosine nearest-centroid decoding forms one prototype per label from the other seven sentences and predicts the exact requested label of the held-out rows. Accuracy is micro-averaged across all 590 retained records. The full centered representation gives 69 correct predictions (11.6949\%); the projected representation gives 56 (9.4915\%). Uniform 80-way chance is 1.25\%, and the empirical permutation reference accounts for the observed within-sentence label inventory.

The fitted basis uses all eight calibration sentences. Each held-out sentence is also centered from its own unlabeled bank of retained controls. If $\mathcal X$ denotes the complete lexical bank, the recorded fold evaluates
\begin{equation}
T_x=T\big(B(\mathcal X),\,p_{a,-x},\,z_{x,a}\big),
\end{equation}
rather than a pipeline with $B(\mathcal X\setminus\{x\})$. Thus the test measures content-held-out label centroids within a fixed, transductively preprocessed calibration representation~\cite{probecontrols}. It does not estimate inductive deployment on a single new utterance without a matched control bank. This boundary applies to the meaning of the accuracy, while the fixed map remains appropriate for the conditional label-permutation comparison.

For alignment, examples of each label are normalized and compared across distinct contents. The mean uses 1,915 dependent content pairs, rather than 1,915 independent utterances. Centered full K/V has mean cosine 0.151710, median 0.1387, and 5th--95th percentiles $[-0.1612,0.5331]$. Projected mean cosine is 0.285311. The increase is a factor of 1.8806; it coexists with a 2.2034-percentage-point decrease in exact-label accuracy.

\subsection{Conditional permutation and compactness}
Each null draw independently permutes labels within every sentence while keeping features, sentence membership, and within-sentence label counts fixed. This destroys consistent request-to-feature correspondence across sentences. Applying the same permutation to all sentences would only rename classes and would not provide the intended null. For $B=1000$ draws, the plus-one estimate is~\cite{permutation}
\begin{equation}
\widehat p=\frac{1+\sum_{b=1}^{B}\boldsymbol1[T_b\ge T_{\rm obs}]}{B+1}.
\end{equation}
Neither statistic has a null draw at least as large as its observed value, giving $\widehat p=1/1001\approx0.000999$ for both. The rounded abstract value is 0.001. This is the finite experiment's resolution; it is not an exact analytic tail probability.

\begin{apptable}
\caption{\textbf{The complete projected-geometry null summary.} Quantiles refer to 1,000 conditional permutations; LOCO is exact 80-way centroid accuracy in percent.}
\label{tab:nullfull}
\setlength{\tabcolsep}{3pt}
\begin{tabular}{@{}lrr@{}}\toprule
Statistic & LOCO (\%) & Mean cosine\\\midrule
Observed & 9.4915 & 0.28531\\
Null mean & 1.2468 & 0.01703\\
Null median & 1.1864 & 0.01647\\
Null 5th percentile & 0.5085 & 0.00443\\
Null 95th percentile & 2.0339 & 0.03117\\
Monte Carlo $p$ & 0.000999 & 0.000999\\
Permutations & 1,000 & 1,000\\\bottomrule
\end{tabular}
\end{apptable}

The raw rank-16 SVD captures 61.7621\% of residual variance. This value belongs to the raw decomposition, before content residualization. The historical summary's \texttt{rank\_95\_energy} field is capped at the saved maximum rank of 32. Its returned value of 32 cannot distinguish attainment of 95\% at that rank from a higher uncapped rank. The reported compactness claim therefore uses the verified rank-16 energy, without claiming 95\% compression to 32 dimensions or the same variance fraction for the final projected basis.

\subsection{Why alignment can rise while decoding falls}
An orthogonal projection changes both the cosine numerator and its normalization. For $u=u_\parallel+u_\perp$ and $v=v_\parallel+v_\perp$,
\begin{equation}
\begin{aligned}
\cos(u,v)&=\frac{u_\parallel^\top v_\parallel+u_\perp^\top v_\perp}{D_uD_v},\\
D_w&=\sqrt{\|w_\parallel\|^2+\|w_\perp\|^2},\quad w\in\{u,v\}.
\end{aligned}
\end{equation}
Removing content-dependent variation can align the same label across contents while also removing a distinction between two labels. A centered mathematical example illustrates the compatibility. Let four labels be $(f,b)\in\{-1,+1\}^2$, let two contents be $s\in\{-1,+1\}$, and define
\begin{equation}
\begin{aligned}
x_{f,b,s}&=(f,\ sLf,\ b\epsilon),\qquad 0<L<1,\quad\epsilon>0,\\
P&=\operatorname{diag}(1,0,0).
\end{aligned}
\end{equation}
The four labels average to zero at each content. Their same-label cross-content cosine is $(1-L^2+\epsilon^2)/(1+L^2+\epsilon^2)$ before projection and one afterwards. Yet $Px_{f,+1,s}=Px_{f,-1,s}$, so projection loses $b$. Training cosine centroids on one content and testing on the other, the correct full-space dot product exceeds the same-$f$, opposite-$b$ alternative by $2\epsilon^2$, and the opposite-$f$, same-$b$ alternative by $2(1-L^2)$. Both are positive. Full-space accuracy is one; a fixed tie-breaking rule in the projected space gives 50\% on balanced labels.

This example is an analytic illustration, not measured GLM geometry. It explains why the observed alignment gain and decoding loss are consistent. The aggregate does not identify which GLM distinctions were discarded, support a paired correctness test, or measure an information-retention percentage.

\section{Coordinate Operations and the Generation Interface}
\label{app:operators}

\subsection{The shared projector is the operative object}
For fixed prototypes, a static update depends on $P_\perp=B_\perp B_\perp^\top$, its scale, and its injection sites. The individual axes are not uniquely identifiable. For orthogonal $Q\in\R^{r\times r}$, $B_\perp Q$ with coordinates $Q^\top c_a$ reconstructs the same $d_a$. The following identities hold exactly:
\begin{align}
B_\perp(c_a+c_b)&=P_\perp(p_a+p_b),\\
\begin{split}
B_\perp((1-\lambda)c_a+\lambda c_b)
&=(1-\lambda)d_a+\lambda d_b,
\end{split}\\
\|B_\perp c\|_2&=\|c\|_2.
\end{align}
They explain the common interface: attribute addition is projection of a summed prototype, temporal interpolation is interpolation of reconstructed directions, and coordinate normalization controls their global Euclidean norm. They are linear-algebra properties; perceptual composition and timed acoustic realization remain empirical questions.

\subsection{Matched controls and layer masking}
Prompt-only retains the same model and request without injection. A Gaussian unit vector $u=g/\|g\|_2$, $g\sim\mathcal N(0,I)$, supplies the direction control. For selected-layer target chunks $d_t$, the random update is $\alpha\|d_t\|_2u$. Dynamic items hold $u$ fixed while its magnitude follows the target schedule. This matches the concatenated K/V norm after layer selection. It does not separately match every layer, the K versus V norm, the attention-logit effect, or the downstream response.

For composition, $v_\mathcal C=\sum_{a\in\mathcal C}p_a$ and $d_\mathcal C=P_\perp v_\mathcal C$. The full-space control is
\begin{equation}
\bar d_\mathcal C^{\rm full}=\frac{\|d_\mathcal C\|_2}{\|v_\mathcal C\|_2}v_\mathcal C,
\end{equation}
when both norms are nonzero. Undefined zero-norm matches are rejected. In the all-layer comparison this holds the intervention norm fixed and changes its direction. The mean and normalized-sum variants change magnitude as well as construction: with $m$ constituents, the mean is $1/m$ times the sum. Their score differences do not alone identify nonlinear composition.

Layer groups are early $\{16,20\}$, middle $\{24,28\}$, and late $\{32,36\}$. These names refer to the sampled groups. Masking removes the other chunks without redistributing their energy, so the ablation compares practical injection configurations. It is not an energy-matched causal localization study. Similarly, rank changes both available directions and their norms under the recorded scaling.

\subsection{Temporal schedules}
For a requested transition $a\to b$, $c_t=(1-\lambda_t)c_a+\lambda_tc_b$. Gradual interpolation uses $\lambda_t=\min(t/191,1)$; the sudden schedule switches from zero to one at forward pass 96. Start-only, end-only, and midpoint-only use constant coefficients zero, one, and one-half. The endpoint is held after the 192-pass schedule. Figure~\ref{fig:appendix_schedule} shows the prescribed coefficients.

\begin{appfigure*}
\includegraphics[width=.84\textwidth]{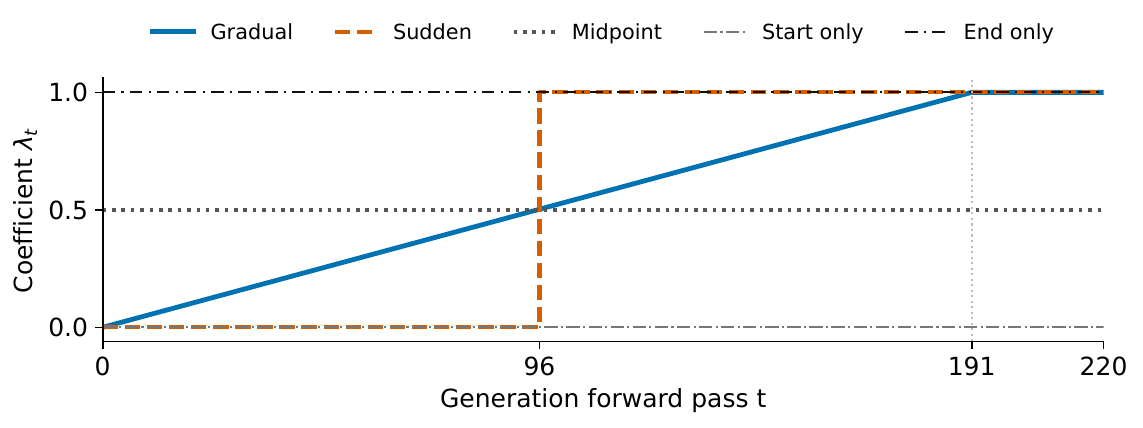}
\caption{\textbf{Prescribed coordinate schedules, not measured acoustic trajectories.} The x-axis counts generation forward passes. Gradual interpolation reaches one at pass 191; the sudden switch occurs at 96; constant controls provide the start, end, and midpoint comparisons. Interleaved text/audio tokens make this index different from waveform time.}
\label{fig:appendix_schedule}
\end{appfigure*}

Early termination can prevent a trajectory from reaching its endpoint, and a forward-pass midpoint need not be the acoustic midpoint. A preference response to the requested transition therefore does not verify its waveform timing. The constant-coordinate controls test whether the implemented path contributes beyond one fixed point under the same generation interface.

\subsection{Representation and intervention have distinct dependencies}
Replay features summarize listening to an utterance; the update acts while producing new text/audio tokens. Their model weights are shared, but prompts, token roles, and causal states differ. Locally, an output statistic $f(q)$ responds approximately as $f(q+\delta)-f(q)\approx J_f(q)\delta$. The covariance used to fit $B_\perp$ does not identify this Jacobian. A decodable direction can consequently have a weak or unintended generation effect~\cite{amnesic,targeted_erasure}. This is a distinction between estimands, not a measured diagnosis of the aggregate controller result.

For an ordinary attention query, adding a common post-position-transform key offset to every unmasked source position shifts all logits equally and leaves their softmax unchanged. A common value offset instead adds that offset to the weighted value sum. These identities require uniform offsets at the stated locations. A pre-attention fused-QKV hook, positional transformations, selective token updates, and caching need not satisfy those assumptions. They clarify why K/V intervention details matter without implying that the recorded hook is inert.

\section{Evaluation Protocol and Complete Held-Out Results}
\label{app:selection}

\subsection{Coverage, development, and selected scales}
Evaluation uses catalog-covered English SpeechParaling-Bench requests and its waveform evaluation path~\cite{speechparaling}. Static requires one covered control, composition requires every constituent, and dynamic requires both endpoints. Coverage filtering precedes development partitioning and does not use judge scores. Positions $0,5,10,\ldots$ in the eligible list form development; all remaining positions are held out. These are task subsets with a matched prompt-only reference, rather than an official overall multilingual leaderboard run. Situational adaptation is outside this evaluation.

The development rule maximizes preference over $\alpha\in\{0.25,0.5,1.0,1.5\}$ among complete candidates whose mean emitted-text WER degradation is at most 0.02. Ties favor the smaller scale. Table~\ref{tab:dev} preserves the full available grid. Static selects 0.5. Its 1.5 candidate is excluded because only 19 of 42 baseline items are present. No composition candidate passes the fidelity gate. Dynamic selects 1.5, but that held-out arm stops at 16 of 60 items; the complete 0.25 arm is reported as exploratory. Neither incomplete selection nor a failed gate is converted into a formal test result.

\begin{apptable*}
\caption{\textbf{Development scale selection.} Preference and nominal 95\% intervals are against prompt-only. $\Delta$WER is an absolute change in emitted-text WER; the tolerance is $+0.02$.}
\label{tab:dev}
\setlength{\tabcolsep}{4pt}
\begin{tabular}{@{}llrrrrl@{}}\toprule
Task & $\alpha$ & Items & Preference & 95\% CI & $\Delta$WER & Status\\\midrule
Static &0.25&42&53.57&[40.48,65.48]&$+0.03335$&Fidelity gate not met\\
Static &0.5&42&46.43&[33.33,59.52]&$+0.00413$&Selected\\
Static &1.0&42&42.86&[29.76,55.95]&$-0.06605$&Eligible\\
Static &1.5&19/42&---&---&---&Incomplete; excluded\\\midrule
Composition &0.25&10&70.83&[55.00,85.83]&$+0.12000$&Fidelity gate not met\\
Composition &0.5&10&70.00&[47.50,90.00]&$+0.12000$&Fidelity gate not met\\
Composition &1.0&10&74.17&[58.33,88.33]&$+0.12000$&Fidelity gate not met\\
Composition &1.5&10&75.83&[53.33,95.00]&$+0.12000$&Fidelity gate not met\\\midrule
Dynamic &0.25&15&46.67&[23.33,70.00]&$-0.00562$&Eligible\\
Dynamic &0.5&15&50.00&[26.67,73.33]&$-0.01022$&Eligible\\
Dynamic &1.0&15&43.33&[20.00,66.67]&$+0.00106$&Eligible\\
Dynamic &1.5&15&53.33&[30.00,76.67]&$-0.01172$&Selected\\\bottomrule
\end{tabular}
\end{apptable*}

\subsection{Preference, fidelity, and uncertainty}
For item $i$ and judged dimension $j$, a win, tie, or loss receives $w_{ij}\in\{1,\tfrac12,0\}$. With $m_i$ judged dimensions, $s_i=m_i^{-1}\sum_jw_{ij}$ and reported preference is $100n^{-1}\sum_is_i$. Every item receives equal weight even when it supplies several dimension decisions. A value of 50 denotes parity and is not a percentage of strict wins. For example, the 38 composition items supply 88 dimension decisions; item-level analysis uses 38 scores.

Nominal 95\% preference intervals use 10,000 item-level bootstrap resamples~\cite{significance}. The balanced static extension reports its recorded hierarchical interval for reused item/control comparisons. All intervals below are reproduced from the supplied summaries, rather than recalculated from unavailable raw judgments. Configuration intervals test preference relative to the stated comparator; ordering two estimates does not supply a paired significance test between them.

Text fidelity is recorded as WER between the intended words and the model's generated text channel. The public summaries do not provide independent waveform ASR, a listening study, or a timing measurement. The local evaluator wrapper uses upstream English static, composition, and dynamic modules. Its saved paths do not pin the run-level judge checkpoint, service version, and upstream revision. SpeechParaling's published evaluator describes a model-based pairwise judge; recovering the exact evaluator used in these runs requires the original metadata rather than assuming the current upstream version.

\subsection{Task-level results and their comparison targets}
Table~\ref{tab:heldfull} reports the full-task comparisons, including the direction controls. Static at the selected scale scores 55.36 against prompt-only, while its direction control scores 55.95. Those are two baseline comparisons; their difference is not a head-to-head win rate. The balanced direct comparison in Appendix~\ref{app:static_power} is the relevant available test of direction specificity.

\begin{apptable*}
\caption{\textbf{Complete reported task-level preference comparisons.} All entries use item-averaged scores (50 = parity). ``Selected'' identifies the completed development-selected static arm. Composition has no admissible development scale; the complete dynamic arm uses exploratory $\alpha=0.25$.}
\label{tab:heldfull}
\setlength{\tabcolsep}{5pt}
\begin{tabular}{@{}llrrrl@{}}\toprule
Task & Candidate versus comparator & Items & Preference & 95\% CI & Status\\\midrule
Static & ParaGeo vs prompt-only &168&55.36&[49.11,61.61]&Selected, $\alpha=0.5$\\
Static & Random vs prompt-only &168&55.95&[49.40,62.50]&Direction control\\\midrule
Composition & ParaGeo vs prompt-only &38&45.39&[36.18,54.82]&Exploratory\\
Composition & Full-space vs prompt-only &38&48.03&[37.94,58.33]&Exploratory\\
Composition & ParaGeo vs full-space &38&42.98&[34.87,51.10]&Exploratory direct\\\midrule
Dynamic & ParaGeo vs prompt-only &60&51.67&[40.00,63.33]&Exploratory\\
Dynamic & Random vs prompt-only &60&46.67&[35.00,58.33]&Exploratory control\\\bottomrule
\end{tabular}
\end{apptable*}

\begin{apptable}
\caption{\textbf{Complete ParaGeo text-channel fidelity.} WER and paired changes are absolute fractions. A change of $+0.02$ is two percentage points, rather than a two-percent relative increase.}
\label{tab:held_fidelity}
\setlength{\tabcolsep}{3pt}
\begin{tabular}{@{}lrrrr@{}}\toprule
Task & Items & ParaGeo & Prompt-only & $\Delta$WER\\\midrule
Static &168&0.30526&0.31254&$-0.00729$\\
Composition &38&0.45671&0.38789&$+0.06883$\\
Dynamic &60&0.26093&0.23537&$+0.02556$\\\bottomrule
\end{tabular}
\end{apptable}

The recorded controller acceptance rule additionally requires a five-point gain for static/composition or an eight-point gain for dynamic, together with fidelity and random-control conditions. No complete task satisfies this combined rule. This retrospective reporting preserves the rule in the source record without treating it as a newly passed prospective gate. The favorable configurations below characterize responses of the fitted controller rather than replace that task-level assessment.

\section{Complete Rank, Projection, and Layer Study}
\label{app:ablations}

\subsection{A fixed diagnostic subset}
Each task's configuration study uses the first 24 eligible held-out items in benchmark-index order. The subset is fixed but can have a different family mix from the full task. Its purpose is to inspect rank, projection, injection location, and schedule under a consistent selection. It is not newly sampled confirmation of a configuration chosen from these outcomes. Tables~\ref{tab:abl_static}--\ref{tab:abl_dynamic} retain every recorded configuration, its nominal preference interval, and its text-fidelity change.

The main condition uses all six sampled layers, rank 16, the content-associated projection, summed coordinates for composition, and the request-dependent temporal schedule. The no-projection condition uses the raw basis. Rank variants use the same stored decomposition at ranks 4, 8, and 32. The static rank-4 preference has 23 judged inputs, while its WER summary has 24; the mismatch is explicit and constrains comparison with the other rows.

\begin{apptable}
\caption{\textbf{Complete static configuration results.} The first 24 eligible held-out inputs form the fixed diagnostic subset. Preference is against prompt-only (50 = parity); intervals are nominal 95\% item-bootstrap CIs. $\Delta$WER is an absolute text-channel change. Static rank-4 preference has 23 judged inputs; its WER summary has 24.}
\label{tab:abl_static}
\setlength{\tabcolsep}{2pt}
\begin{tabular}{@{}lrrrr@{}}\toprule
Configuration & $n$ & Pref. & 95\% CI & $\Delta$WER\\\midrule
Main (all; rank 16) & 24 & 60.42 & [45.83,72.92] & $-0.0528$\\
No content projection & 24 & 56.25 & [39.58,70.83] & $-0.0528$\\
Rank 4 & 23 & 52.17 & [36.96,65.22] & $-0.0403$\\
Rank 8 & 24 & 58.33 & [41.67,75.00] & $-0.0528$\\
Rank 32 & 24 & 56.25 & [37.50,72.92] & $+0.0065$\\
Early layers & 24 & 52.08 & [37.50,66.67] & $-0.0528$\\
Middle layers & 24 & 64.58 & [50.00,79.17] & $-0.0278$\\
Late layers & 24 & 58.33 & [45.83,70.83] & $-0.0153$\\
\bottomrule\end{tabular}
\end{apptable}

\begin{apptable}
\caption{\textbf{Complete composition configuration results.} The first 24 eligible held-out inputs form the fixed diagnostic subset. Preference is against prompt-only (50 = parity); intervals are nominal 95\% item-bootstrap CIs. $\Delta$WER is an absolute text-channel change.}
\label{tab:abl_composition}
\setlength{\tabcolsep}{2pt}
\begin{tabular}{@{}lrrrr@{}}\toprule
Configuration & $n$ & Pref. & 95\% CI & $\Delta$WER\\\midrule
Main (all; rank 16) & 24 & 66.67 & [54.17,78.12] & $+0.1250$\\
No content projection & 24 & 48.96 & [35.42,61.46] & $-0.0333$\\
Rank 4 & 24 & 53.12 & [37.50,68.75] & $-0.0333$\\
Rank 8 & 24 & 50.00 & [38.54,61.46] & $-0.0333$\\
Rank 32 & 24 & 56.25 & [43.75,68.75] & $+0.1125$\\
Early layers & 24 & 42.71 & [30.21,55.21] & $-0.0542$\\
Middle layers & 24 & 54.17 & [44.79,64.58] & $+0.0139$\\
Late layers & 24 & 46.88 & [39.58,54.17] & $-0.0556$\\
Mean sum & 24 & 59.38 & [50.00,68.75] & $+0.1250$\\
Normalized sum & 24 & 43.75 & [31.25,56.25] & $-0.0347$\\
\bottomrule\end{tabular}
\end{apptable}

\begin{apptable}
\caption{\textbf{Complete dynamic configuration results.} The first 24 eligible held-out inputs form the fixed diagnostic subset. Preference is against prompt-only (50 = parity); intervals are nominal 95\% item-bootstrap CIs. $\Delta$WER is an absolute text-channel change.}
\label{tab:abl_dynamic}
\setlength{\tabcolsep}{2pt}
\begin{tabular}{@{}lrrrr@{}}\toprule
Configuration & $n$ & Pref. & 95\% CI & $\Delta$WER\\\midrule
Main (all; rank 16) & 24 & 54.17 & [37.50,70.83] & $+0.0055$\\
No content projection & 24 & 43.75 & [27.08,60.42] & $+0.0021$\\
Rank 4 & 24 & 58.33 & [41.67,75.00] & $+0.0000$\\
Rank 8 & 24 & 54.17 & [37.50,70.83] & $+0.0108$\\
Rank 32 & 24 & 47.92 & [31.25,64.58] & $+0.0021$\\
Early layers & 24 & 60.42 & [45.83,75.00] & $+0.0160$\\
Middle layers & 24 & 68.75 & [56.25,81.25] & $-0.0060$\\
Late layers & 24 & 60.42 & [45.83,75.00] & $+0.0000$\\
Start only & 24 & 56.25 & [39.58,70.83] & $+0.0010$\\
End only & 24 & 47.92 & [31.25,66.67] & $+0.0327$\\
Midpoint only & 24 & 45.83 & [29.17,62.50] & $+0.0000$\\
\bottomrule\end{tabular}
\end{apptable}

\subsection{The favorable layer setting and the rank response}
The middle group has the highest layer-group point estimate for static and dynamic requests. Static scores 64.58 [50.00,79.17], with $\Delta$WER $=-0.0278$. Dynamic scores 68.75 [56.25,81.25], with $\Delta$WER $=-0.0060$. The dynamic result is a positive response relative to prompt-only at that operating point, with no mean text-channel fidelity loss on the subset. Its interval is nominal and unadjusted for inspecting multiple configurations; the record does not contain a direct, energy-matched test against the all-layer condition.

The composition result supplies a different operating point. All-layer rank-16 addition scores 66.67 [54.17,78.12] on the 24-input subset, but increases text WER by 0.1250. Middle-layer composition scores 54.17 with a smaller change of 0.0139. Thus the largest preference in this subset is accompanied by a material fidelity cost. This preserves the positive measured response and explains why it cannot establish joint compositional improvement.

Rank response is nonmonotone. Static preferences at ranks 4, 8, 16, and 32 are 52.17, 58.33, 60.42, and 56.25; dynamic values are 58.33, 54.17, 54.17, and 47.92. Retaining more raw variance therefore does not monotonically improve the recorded intervention score. These are practical rank settings with different reconstructed vectors and norms. They do not isolate the causal effect of dimension count or measure a correlation between layerwise decoding and steering.

Removing content projection gives static preference 56.25 and dynamic preference 43.75, compared with 60.42 and 54.17 for the projected main setting. On composition, projection changes both preference and text disturbance. The comparison supports treating the projection as a consequential design choice; its generation results do not validate complete semantic disentanglement. Earlier injections also change later inputs, so all-layer intervention can reinforce or counteract later perturbations. The reported layer profiles reveal the dependence without identifying its mechanism.

\begin{appfigure*}
\includegraphics[width=.82\textwidth]{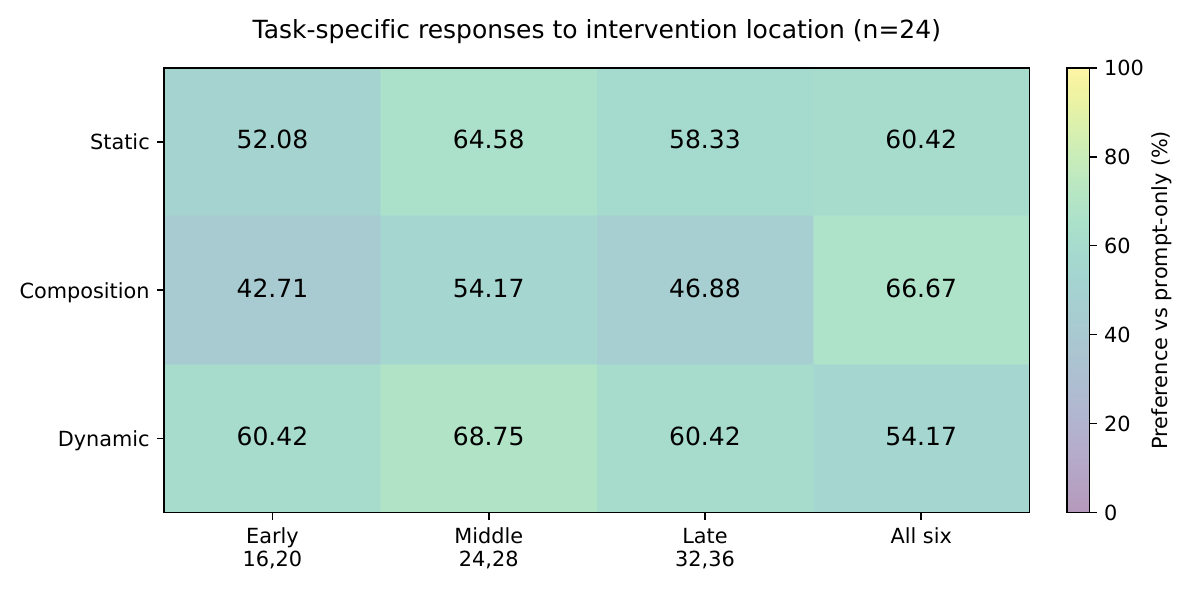}
\caption{\textbf{The layer response depends on the task.} All four sampled-layer configurations are shown for all three 24-input diagnostic tasks. Cells are preference estimates against prompt-only; their recorded nominal intervals appear in Tables~\ref{tab:abl_static}--\ref{tab:abl_dynamic}. Middle-layer injection has the largest static and dynamic estimates; all-layer composition has the largest composition estimate and a substantial text-fidelity cost (Table~\ref{tab:abl_composition}). The matrix is a response profile, not a corrected test of pairwise layer superiority.}
\label{fig:appendix_layer_matrix}
\end{appfigure*}

\section{Balanced Static Responses Across Attributes and Controls}
\label{app:static_power}

\subsection{A direct comparison at a common setting}
The expanded study tests all nine Attitude and nine Cognitive State controls with ten inputs per label: 180 unique inputs. Each ParaGeo output is compared with five random-direction conditions, giving 900 reused candidate--comparator comparisons. S1--S5 are anonymous identifiers for the direction-control conditions. These are control orientations, rather than independently trained ParaGeo models or five independent speech-generation replications. Tables~\ref{tab:powerattrs} and \ref{tab:powerseeds} give the full attribute and control summaries.

\begin{apptable*}
\caption{\textbf{All 18 attributes against all five random-direction conditions.} Each row uses ten unique inputs, with the same ParaGeo outputs reused across S1--S5. The final columns are descriptive mean, minimum, and count strictly above 50. Labels retain the source order; no unfavorable row is excluded.}
\label{tab:powerattrs}
\setlength{\tabcolsep}{7pt}
\begin{tabular}{@{}lrrrrrrrr@{}}\toprule
Attribute & S1 & S2 & S3 & S4 & S5 & Mean & Min. & $>50$\\\midrule
Cold & 35 & 55 & 10 & 50 & 55 & 41 & 10 & 2/5\\
Contemptuous & 35 & 30 & 35 & 50 & 20 & 34 & 20 & 0/5\\
Enthusiastic & 55 & 40 & 50 & 70 & 80 & 59 & 40 & 3/5\\
Perfunctory & 40 & 30 & 45 & 30 & 45 & 38 & 30 & 0/5\\
Polite & 60 & 75 & 55 & 55 & 75 & 64 & 55 & 5/5\\
Rude & 40 & 45 & 45 & 60 & 65 & 51 & 40 & 2/5\\
Sarcastic & 60 & 65 & 35 & 80 & 70 & 62 & 35 & 4/5\\
Sincere & 40 & 55 & 55 & 35 & 55 & 48 & 35 & 3/5\\
Teasing & 50 & 65 & 70 & 65 & 75 & 65 & 50 & 4/5\\
Anxious & 70 & 20 & 50 & 65 & 50 & 51 & 20 & 2/5\\
Confident & 65 & 60 & 50 & 35 & 65 & 55 & 35 & 3/5\\
Confused & 35 & 55 & 65 & 55 & 60 & 54 & 35 & 4/5\\
Curious & 45 & 45 & 30 & 30 & 40 & 38 & 30 & 0/5\\
Doubting & 40 & 60 & 65 & 50 & 60 & 55 & 40 & 3/5\\
Helpless & 60 & 60 & 60 & 45 & 25 & 50 & 25 & 3/5\\
Hesitant & 50 & 60 & 20 & 40 & 45 & 43 & 20 & 1/5\\
Nervous & 40 & 65 & 75 & 15 & 45 & 48 & 15 & 2/5\\
Tired & 70 & 70 & 70 & 80 & 65 & 71 & 65 & 5/5\\
\bottomrule\end{tabular}
\end{apptable*}

\subsection{Which responses persist across the inspected controls?}
Tired has preference values [70,70,70,80,65], mean 71, and minimum 65; polite has [60,75,55,55,75], mean 64, and minimum 55. Each exceeds 50 in all five columns. Teasing has mean 65, with four columns above parity and one tie. Sarcastic has mean 62, with four above-parity columns but a minimum of 35. The minimum and positive-column counts distinguish a favorable mean from a pattern that persists across the inspected control orientations. They remain descriptions of the same small input groups.

The complete matrix also contains adverse responses. Contemptuous averages 34, perfunctory and curious 38, and cold 41. Some labels are strongly control-dependent: nervous ranges from 15 to 75, while anxious ranges from 20 to 70. These entries are essential to understanding why a favorable response for one requested style does not extend to the whole family. The sorted heatmap in main-text Fig.~\ref{fig:attribute_response} presents the same values; the native table here retains the original label order for exact lookup.

A descriptive two-way sum-of-squares decomposition of the 18-by-5 cell means assigns 40.15\% of cell variation to attribute means, 2.19\% to control-column means, and 57.66\% to the remaining interaction plus unseparated sampling/judge residual. This is a summary of the observed matrix, not inferential ANOVA or a causal variance decomposition. It emphasizes that response depends substantially on the requested attribute and its interaction with the comparison setting, rather than on a single pooled mean.

\begin{apptable}
\caption{\textbf{Direct ParaGeo preference for each direction-control condition.} Every row reuses the same 180 inputs. The source intervals and strict win/tie/loss counts are preserved. Control orientations do not provide independent ParaGeo runs.}
\label{tab:powerseeds}
\setlength{\tabcolsep}{3pt}
\begin{tabular}{@{}lrrr@{}}\toprule
Control & Pref. & 95\% CI & W/T/L\\\midrule
S1 & 49.44 & [41.11,58.06] & 75/28/77\\
S2 & 53.06 & [43.61,62.22] & 81/29/70\\
S3 & 49.17 & [38.89,59.17] & 79/19/82\\
S4 & 50.56 & [40.56,60.28] & 74/34/72\\
S5 & 55.28 & [45.56,64.72] & 85/29/66\\
\bottomrule\end{tabular}
\end{apptable}

\subsection{The aggregate result and fidelity}
Pooling all labels and controls gives preference 51.50 with the recorded hierarchical interval [45.39,57.56]. Three of five control-specific estimates exceed 50. The source claim rule requires both a pooled lower bound above 50 and positive point estimates in at least four columns, so this record does not pass that rule. It also does not establish equivalence between learned and random updates; a near-parity estimate with a wide interval answers neither a superiority nor an equivalence question by itself.

Mean ParaGeo text WER is 0.19509. Paired mean WER differences relative to S1--S5 are approximately $+0.00646$, $-0.00572$, $-0.00611$, $-0.00191$, and $-0.00113$. Their small size indicates that this comparison is not accompanied by a large relative text-channel penalty. The remaining uncertainty concerns direction specificity and attribute response at this setting. Similar scores may reflect nonspecific acoustic changes, generation stochasticity, judge variation, or their combination; the aggregate does not separate those explanations.

Tired and polite were highlighted after the complete matrix was inspected. Their consistency is useful exploratory evidence for attribute-dependent controllability, but shared outputs, ten-input label groups, and subgroup selection prevent treating five positive columns as independent confirmation. This distinction preserves the practical value of the observed patterns while keeping the aggregate test visible.

\section{Composition and Dynamic Family Profiles}
\label{app:composition_details}

\subsection{What low-rank composition changes}
The 38-input composition comparison holds the summed prototype construction and concatenated intervention norm fixed while comparing projected and full-space directions (Appendix~\ref{app:operators}). Text WER is 0.45671 with projection and 0.50280 in full space, a descriptive reduction of 4.61 percentage points. Prompt-only WER is 0.38789, so both interventions still disturb the text channel relative to that reference. No paired WER interval is supplied for the projected--full-space difference.

\begin{apptable*}
\caption{\textbf{Composition control and fidelity on the same 38 inputs.} Preference is against prompt-only; both interventions use exploratory $\alpha=0.25$. The final row is the direct preference for projection over full-space addition. WER is in percent here, for direct comparison with the percentage-point changes.}
\label{tab:composition_tradeoff}
\setlength{\tabcolsep}{6pt}
\begin{tabular}{@{}lrrrr@{}}\toprule
Comparison & Preference & 95\% CI & Text WER (\%) & $\Delta$WER (pp)\\\midrule
Prompt-only &---&---&38.79&0.00\\
Projected sum vs prompt-only &45.39&[36.18,54.82]&45.67&$+6.88$\\
Full-space sum vs prompt-only &48.03&[37.94,58.33]&50.28&$+11.49$\\
Projected vs full-space &42.98&[34.87,51.10]&---&$-4.61$\\\bottomrule
\end{tabular}
\end{apptable*}

Direct preference for the projected direction is 42.98 [34.87,51.10]. Of the 38 item-averaged scores, eight are above parity, 17 equal, and 13 below. Their 88 dimension decisions comprise ten wins, 56 ties, and 22 losses. The denominators refer to different units and are both retained. No reported family in the direct comparison exceeds 50. Projection therefore yields less text disturbance at this matched operating point, accompanied by a lower control-preference estimate; it is not demonstrated compositional superiority.

The 24-input configuration study in Appendix~\ref{app:ablations} is a different subset. Its all-layer preference of 66.67 cannot replace the complete 38-input estimate of 45.39. Together, the two results show that favorable local preference and complete-task performance can differ, and that composition scale remains constrained by text fidelity. The mean and normalized-sum ablations help characterize that dependence, but their changed magnitude prevents a pure test of semantic compositionality.

\subsection{All static and dynamic family outcomes}
Table~\ref{tab:families} reports every covered static family and every family with dynamic inputs, including the random-versus-prompt-only references. The columns use separate baseline comparisons. They cannot be subtracted to obtain direct ParaGeo-versus-random preference.

\begin{apptable*}
\caption{\textbf{Every reported static and dynamic family.} Both preference columns compare separately with prompt-only (50 = parity); they are not a head-to-head comparison. Static uses selected $\alpha=0.5$ and dynamic uses exploratory $\alpha=0.25$. Family-specific intervals are unavailable.}
\label{tab:families}
\setlength{\tabcolsep}{10pt}
\begin{tabular}{@{}llrrr@{}}\toprule
Task & Family & Inputs & ParaGeo & Random control\\\midrule
Static & Age & 9 & 66.67 & 50.00\\
Static & Attitude & 22 & 40.91 & 56.82\\
Static & Cognitive State & 22 & 45.45 & 63.64\\
Static & Emotion & 17 & 64.71 & 64.71\\
Static & Non-Linguistic Vocalizations & 19 & 63.16 & 63.16\\
Static & Pace & 12 & 50.00 & 58.33\\
Static & Pause & 2 & 25.00 & 50.00\\
Static & Pitch & 12 & 54.17 & 37.50\\
Static & Rhythm & 17 & 52.94 & 64.71\\
Static & Stress & 9 & 55.56 & 33.33\\
Static & Timbre & 15 & 66.67 & 46.67\\
Static & Volume & 12 & 66.67 & 54.17\\\midrule
Dynamic & Emotion & 12 & 54.17 & 62.50\\
Dynamic & Pace & 12 & 75.00 & 62.50\\
Dynamic & Pitch & 12 & 45.83 & 62.50\\
Dynamic & Rhythm & 12 & 37.50 & 16.67\\
Dynamic & Timbre & 4 & 62.50 & 37.50\\
Dynamic & Volume & 8 & 37.50 & 25.00\\
\bottomrule\end{tabular}
\end{apptable*}

The static profile contains point estimates of 66.67 for age, timbre, and volume, 64.71 for emotion, and 63.16 for non-linguistic vocalizations. Attitude and Cognitive State are below parity in this full static set. The balanced extension uses different inputs and a direct random comparison, so its favorable individual labels answer a different question from those family-level baseline scores.

Dynamic pace has preference 75 over 12 inputs at the common exploratory scale, without family-specific tuning. Emotion scores 54.17, timbre 62.50 over four inputs, pitch 45.83, and rhythm and volume 37.50. The complete dynamic score of 51.67 consequently averages contrasting family responses. These are small descriptive groups without family-specific intervals or corrected subgroup tests. The pace estimate supplies a concrete positive response, while the full family table prevents it from standing in for all temporal controls.

On the 24-input schedule subset, the varying main coordinate scores 54.17, while a constant start coordinate scores 56.25, a constant end 47.92, and a constant midpoint 45.83. The record therefore does not establish an aggregate advantage of the prescribed temporal path over its constant-coordinate controls. It also does not contain measured pitch/energy contours or transition-onset annotations. Such measurements would be needed to distinguish endpoint selection from successful timing of a transition.

\section{Reproducibility, Related Methods, and Scope}
\label{app:reproducibility}

\subsection{Execution settings and artifact identity}
The settings in Table~\ref{tab:settings} apply to the evaluated frozen-backbone snapshot. All empirical summaries originate from repository commit \texttt{b71c6cc5\allowbreak a1faf51f\allowbreak 3e952450\allowbreak 68b4c9c9\allowbreak 241f8c0e}, specifically the geometry, ICASSP evaluation, and balanced static summaries. The manuscript's public code link and the historical result repository identify different stages of the project; the commit, result paths, and numerical summaries fix the evidence used here.

\begin{apptable}
\caption{\textbf{Recorded execution settings.} Restart equivalence also requires traversal and artifact identity.}
\label{tab:settings}
\setlength{\tabcolsep}{3pt}
\begin{tabular}{@{}>{\raggedright\arraybackslash}p{.36\linewidth}>{\raggedright\arraybackslash}p{.57\linewidth}@{}}\toprule
Setting & Value\\\midrule
Backbone; weights & GLM-4-Voice-9B; int4/NF4\\
Temperature; top-$p$ & 0.8; 0.8\\
Maximum new tokens; waveform rate & 768; 22.05 kHz\\
Selected layers & 16, 20, 24, 28, 32, 36\\
Default rank; evaluated alternatives & 16; 4, 8, 32\\
Content-basis setting & 8 returned columns; rank boundary in Appendix~\ref{app:acquisition}\\
Schedule; sudden switch & 192 forward passes; switch at pass 96\\
Development grid & 0.25, 0.5, 1.0, 1.5\\
Development partition; ablation subset & Every fifth eligible input; first 24 held-out inputs\\
Permutation draws; bootstrap resamples & 1,000; 10,000\\
\bottomrule\end{tabular}
\end{apptable}

Feature records must align by branch ID, and stored layer/pooling metadata determine the K/V chunk order. Generation manifests record item, split, variant, scale, seed, transcript, and waveform identity. Random vectors are consumed from a seeded NumPy generator during item traversal. Resuming with skipped items can change this mapping unless the original traversal and saved manifests are preserved. A seed alone is therefore insufficient to guarantee restart equivalence. These details affect reproduction, even though they do not alter the numerical summaries reported here.

\subsection{What the supplied package can reproduce}
The package provides aggregate geometry and confidence bounds, complete configuration scores, per-family summaries, the 18-by-5 response matrix, and the historical discovery matrices. They support the native tables and summary-level plots. The raw K/V NPZ files, every generated waveform, the full singular spectrum, per-item judge outputs, and all permutation draws are not present in the attached packet. The recorded intervals are retained without fabricating replacements from means. A true 80-by-80 prototype geometry, uncapped energy rank, or measured acoustic trajectory requires the original local artifacts. The local exporter can process those artifacts when available; its presence does not imply it has been run on them.

\subsection{Positioning the contribution against existing controls}
Learned style tokens and reference-conditioned prosody transfer make expressive speech factors reusable~\cite{gst,prosodytransfer}; SpeechSplit and SpeechTokenizer separate speech information through learned bottlenecks or token structure~\cite{speechsplit,speechtokenizer}. ParaGeo studies a post-hoc shared projection in a frozen model's speech-replay K/V representation. This describes the object investigated without asserting that learned style factors or latent arithmetic are new.

Activation addition and contrastive addition already construct inference-time behavior directions~\cite{actadd,caa}, and the linear representation hypothesis connects probing and steering through geometry~\cite{linearrep}. In speech, EmoSteer-TTS selects emotion-relevant token contributions; EmoShift learns emotion-specific activation offsets; and CoCoEmo constructs matched emotion contrasts and analyzes steering locations and mixed-emotion behavior~\cite{emosteer,emoshift,cocoemo}. Emo-FiLM targets word-level modulation, DisCo-Speech uses a disentangled codec, and later work examines the geometry of emotion composition~\cite{emofilm,discospeech,emotiongeometry}. Emotion Arithmetic and Task Vector in TTS instead operate in model weight space~\cite{emotionarithmetic,taskvector}.

The present contribution is the shared matched-content description across heterogeneous requested controls and the explicit separation of its geometry diagnostics from its generation-response tests. Same-backbone prompt-only, random, raw-basis, full-space, rank, layer, and schedule controls isolate the recorded operations. Full-space prototype addition is not a reproduced CoCoEmo system: it omits that method's complete extraction, site-selection, renormalization, and evaluation design. None of these controls supplies a performance claim against named published TTS methods. The source packet supports an empirical analysis of a specified representation and interface, rather than a new priority claim for activation steering or composition.

\subsection{Boundaries of the conclusions}
The experiments cover one backbone, English generation requests, a fixed calibration lexical bank, and model-based judgments. The balanced extension reuses one synthetic input voice. Requested labels are not independently annotated realized attributes, and emitted-text matching/WER is not waveform transcription. The globally fitted basis, imperfect content proxy, unadjusted exploratory configuration intervals, reused control comparisons, and missing evaluator version limit the respective generalization and replication claims. These restrictions attach to specific measurements, rather than cancel the observed request-associated structure or favorable response settings.

Age, Cognitive State, and Attitude are generation-request labels. A coordinate for confident tone does not measure a real person's confidence, and an elderly-voice instruction is not a validated age estimator. The record supplies no clinical, demographic, or personality inference. Applications to speaker profiling or consequential decisions would require different evidence and ethical assessment.

The extended results make the practical implication precise: a common coordinate system allows vocal delivery to be compared, combined, and scheduled with fixed speech-model weights; choosing an effective intervention still requires attention to the attribute, injection location, norm, and temporal interface. Training-only basis evaluation, independent acoustic/listener validation, exact evaluator recovery, and fresh confirmation of the favorable settings remain separate evidence obligations.

\end{document}